# Maxwell's Demon and the Information Channel Width of a Black Hole

Liangsuo Shu[1,2], Xiaokang Liu[1], Shiping Jin[1,2,3], Suyi Huang[1]

[1]School of energy and power engineering, [2]Innovation Institute, [3]China-Europe Instiute for Clean and Renewable Energy, Huazhong University of Science & Technology. Wuhan, China.

**Abstract**

Using a new generalized second law of thermodynamics, the information and entropy of a black hole and its accretion disk are analyzed respectively. We find the bound of the information channel width of a black hole, which is determined by the variation rate of the horizon temperature and the mass of the black hole's shell (the accretion disk close to the horizon).

**Introduction**

In the early 1970s, Bekenstein[1] introduced the concept of black-hole entropy and the generalized second law of thermodynamics (GSL, the sum of the black hole entropy and the common entropy in the black-hole exterior never decreases) to solve the entropy problem of black hole. Later, Hawking[2,3] proposed a theoretical argument for the existence of Hawking radiation, which also led to the birth of the black hole information paradox. To solve this paradox, many new concepts have been provided in recent years including black hole complementarity[4–6], the fuzzball[7,8], AMPS firewall[9], the apparent horizon[10], ER=EPR[11], the soft hair[12] and so on.

Although gravity can be described by general relativity presented by Albert Einstein 101 years ago[13], its origin is still not clear. Many researchers have been trying to give an answer to this question using different methods including string theory and loop quantum gravity[14]. Beside these two mainstreams, seeking a thermodynamic interpretation of gravity becomes another possible road simply because Black hole thermodynamics [1-3],[15–17] shows there are some deep relationships between gravity, quantum mechanics and thermodynamics[14]. Along this path, the holographic principle has been found by G.'t Hooft [18] and Susskind[19], and many entropic force theories[20] for gravity have already been put forward following the works of Jacobson[21] and Verlinde[22]. Although entropic gravity cannot be the final verdict regarding the end of gravity as a fundamental force because the same method can also be used to deal with the Coulomb force[23], it can give a new thermodynamic description of gravity at least.

In fact, the history of the discussion regarding the relationship between information and entropy can date back to Maxwell's Demon[24]. After the efforts of many researchers[25–28], we now know that the demon in James Maxwell's thought experiment doesn't violate the second law of thermodynamics because its information processing is accompanied by a neglected entropy increase[29]. Thanks to technological advancements in manipulating small systems, some "demons" have been created in different ingenious experiments[30–32]. The discovery of fluctuation theorems by Evans and his co-workers in 1993[33] and its subsequent generalization[34–36] and application[37] promotes the revival of the study on the relationship between information and entropy. Sagawa and Ueda[38] found a method to deal with fluctuating non-isolated systems (systems that exchanges information with other systems).

This method was used in analyzing the biochemical signal transduction with a feedback loop[39], which was found to have thermodynamic similarity to Maxwell's Demon. We found that their method can also be referenced to analyze the entropy and information of a black hole. A new generalized second law of thermodynamics different from Bekenstein's GSL can be expressed as below:

$$dS - dI \geq 0 \quad (1)$$

where $dS$ is the entropy change of a system, while $dI$ is the change of information between the system and its environment.

Using this new generalized second law of thermodynamics (NGSL) and entropic gravity, we can peel off a black hole from its *shell* (the accretion disk close to the horizon) and analyze the information and entropy of a black hole and its shell respectively. A bound of the information channel width of black hole was found. We also found that if a black hole obeys NGSL in its evolution, a free-falling observer will reach the singularity after an ordinary journey when another outside observer observes that the free-falling one is destroyed.

**Entropic gravity and gravitational information**

A non-gravitating isolated system tends to evolve toward a uniform state with high entropy. Therefore, from the view of a hypothetical alien lifeform living in a world without gravitation, the spontaneous gravitational collapse of a nebula will be an entropy-decreasing process, which violates the second law of thermodynamics just like Maxwell's Demon. Here, the neglected entropy increase because of information processing is the gravitational interaction, which is beyond the recognition capacity of this alien.

The entropic gravity theory developed in recent years enables us to give a quantitative description of the *gravitational information* ($I_g$) of a system. Gravitational information of a marked object in the gravitational field can be defined as,

$$I_{g-m} = \frac{m}{T_{hs}} \quad (2)$$

where $m$ is the energy of the marked object (here, nature units are used in which $G=c=\hbar=k_B=1$), and $T_{hs}$, the generalized Hawking temperature on the holographic screen[23], can be calculated from Unruh's law[40].

$$T_{hs} = \frac{g}{2\pi}e^{\phi} \quad (3)$$

where $g$ is the local gravitational acceleration, and $e^{\phi}$ is the redshift factor[22]. As a thermodynamic theory, entropic gravity can only provide a phenomenological description: any object involved in gravitational interaction (including a singularity) will have the ability to "*sense*" $T$ of the gravitational field through some kinetics mechanisms to "*feel*" its gravity. Kinetics mechanisms (which may be sensing the change of curvature of spacetime, exchanging gravitons, sensing the vibration of strings or by involving other methods) are beyond the scope of this work.

When a marked object moves to the center of the gravitational field, from the view of an observer at infinity, its $I_g$ will decrease and contribute an additional entropy increase ($dS_I$) according to the second law of thermodynamics. Also, the gravity of the marked object is an entropic force of $dS_I$ in entropic gravity theories.

$$dS_I = -dI_{g-m} = -d(\frac{m}{T_{hs}}) \qquad (4)$$

From the view of an observer free falling with the marked object (or the marked object itself), it is an inertial motion without any force as well as entropy increase. Therefore, its $I_g$ is a constant, which can be regarded as the value of $I_{g-m}$ in equation (2) when the object reaches the surface of the gravity source for the convenience of discussion.

When the marked object reaches the horizon of the black hole, the temperature in equation (2) will be the Hawking temperature of the horizon of the black hole ($T_H = 1/8\pi M$), which is also the temperature of the holographic screen of the external environment from the view of the singularity of a black hole because that the gravity is an interaction between two mass. Therefore, the gravitational information of a Schwarzschild black hole with a mass of $M$ is

$$I_{g-M} = \frac{M}{T_H} \qquad (5)$$

**New generalized second law of thermodynamics and black hole**

When a particle (Alice) with mass $m_a$ passes the horizon of a Schwarzschild black hole, its $I_g$ will increase

$$dI_{g-M} = Md(\frac{1}{T_H}) + \frac{dM}{T_H} \qquad (6)$$

where $dM = m_a$, and $d(1/T_H) = 8\pi m_b$. There are two equal terms on the right side of the above equation, the first one ($dI_{g-M-1}$) is the result of the mass at singularity *sensesing* the change of $T_H$ (Here, the sensor is the singularity itself, which is involved in gravitational interaction), while the second one ($dI_{g-M-2}$) is the result of the mass increase of the black hole (in fact, it is the gravitational information carried by Alice).

For a growing black hole ($dM > 0$), it is an information writing process resulting in an entropy decrease. During the same process, the change of the black hole entropy ($dS_M$) once Alice passes the horizon is,

$$dS_M = \frac{dM}{T_H} = \frac{m_a}{T_H} \qquad (7)$$

There are two ways for an evolving black hole to maintain NGSL,

$$dS - dI = \frac{m_a}{T_H} - dI_{g-M-1} = 0 \qquad (8.a)$$

$$dS - dI = \frac{m_a}{T_H} - dI_{g-M-2} = 0 \qquad (8.b)$$

Therefore, any observer can only get one of $dI_{g-M-1}$ and $dI_{g-M-2}$: if you get one information change, the other one will be destroyed. Equation (8.a) and (8.b) can provide a possible solution to the *firewall*.

The AMPS firewall[9], an important hypothesis in solving the information paradox of black hole in recent years, seems to challenge Einstein's equivalence principle. If an observer gets $dI_{g-M-1}$ and obeys equation (8.a), the information carried by Alice will be destroyed by a firewall. However, if an observer (such as Alice herself) gets $dI_{g-M-2}$ and obeys equation (8.b), she cannot get the information about the temperature change of $T_H$ of the horizon of the black hole (from entropic gravity, it is a precondition for Alice to *feel* gravity) and will reach the singularity after an ordinary journey. In this way, both Einstein's equivalence principle and the monogamy of entanglement can be protected. The special relationships between $I_{g-M-1}$ and $I_{g-M-2}$, as equal to each other and not capable of being achieved simultaneously, indicates that they are not independent but entangled with each other. It is not strange if the information on a holographic screen is entangled with the gravitational information carried by the energy in space surrounded by it. This agrees with the holographic principle, which is the cornerstone of entropic gravity.

From the discussion above, it can be found that our solution to firewall can be regarded as an upgraded version of black hole complementarity.

**Shell and information channel width of a black hole**
When Alice passes the horizon, the entropy change of the exterior environment of a black hole will be

$$dS_{M_s} = \frac{dM_s}{T_H} = -\frac{m_a}{T_H} \qquad (9)$$

where $M_s$ is the mass of *the shell of the black hole*, the definition of which will be given later. The change of the gravitational information of this shell is

$$dI_{g-M_s} = M_s d(\frac{1}{T_H}) + \frac{dM_s}{T_H} \qquad (10)$$

The accretion disk close to the horizon can sense the change of $T_H$, which will affect the gravity of the mass in the accretion disk, and contributes the first term on the right side of the above equation. Since the gravity propagation speed is limited, the affected accretion disk while Alice passes the horizon is also limited. This affected accretion disk was defined as *the shell of a black hole*. The gravitational information carrying by Alice contributes the second term of the right side of the above equation.

When Alice passes the horizon and goes into the black hole (or comes out), an outside observer will lose the information about Alice. Assuming this information change is *dI*, the accompanying entropy change will be – *dI*. Applying NGSL to the shell, we can get

$$dS_{M_s} - dI_{g-M_s} - dI \geq 0 \tag{11}$$

Then using equation (9) and (10), we can get

$$dI \leq -M_s d(\frac{1}{T_H}) \tag{12}$$

The above equation gives the bound of the information channel width of a black hole. We can find that the communication capability between a black hole and an outside observer is determined by the variation rate of the horizon temperature and the mass of the black hole's shell.

**Conclusion**

Using NGSL, a method developed in the thermodynamics of information[38], and entropic gravity, we peel off a black hole from its shell and got the bound of the information channel width of a black hole. If NGSL is obeyed in the evolution of a black hole, the information on a holographic screen should be entangled with the gravitational information carried by the energy in space surrounded by it. In this way, a free-falling observer will reach the singularity after an ordinary journey while another outside observer observes that the free-falling one is burned to ash by a firewall.